\documentclass{nature-wfig}
\usepackage{graphicx}
\usepackage{amsmath,amssymb}
\usepackage{url}
\usepackage{chngcntr}
\usepackage{float}
\floatstyle{plaintop}
\restylefloat{table}

\def\gtrsim{\mathrel{\hbox{\rlap{\hbox{\lower5pt\hbox{$\sim$}}}\hbox{$>$}}}}

\newcommand\ccc{$^{13}$C$^{18}$O}
\newcommand\cc{C$^{18}$O}
\newcommand\kms{km s$^{-1}$}
\newcommand\kl{k$\lambda$}

\title{Mass inventory of the giant-planet formation zone in a solar nebula analog}

\author{Ke Zhang$^{1}$, Edwin A. Bergin$^{1}$, Geoffrey A. Blake$^{2}$, L. Ilsedore Cleeves$^{3}$, Kamber R. Schwarz$^{1}$}

\begin{document}

\maketitle

{\small
\begin{affiliations}
 \item Department of Astronomy, University of Michigan, 1085 S University Ave., Ann
Arbor, MI 48109, USA
\item Division of Geological \& Planetary Sciences, MC 150-21, California Institute of Technology, 1200 E California Blvd., Pasadena, CA 91125, USA
\item Harvard-Smithsonian Center for Astrophysics, 60 Garden Street, Cambridge, MA 02138, USA
\end{affiliations}}

\begin{abstract}

The initial mass distribution in the solar nebula is a critical input to planet formation models that seek to reproduce today's Solar System\cite{Pollack96}. Traditionally, constraints on the gas mass distribution are derived from observations of the dust emission from disks\cite{Andrews07, Isella09}, but this approach suffers from large uncertainties in grain growth and gas-to-dust ratio\cite{Andrews07}. On the other hand, previous observations of gas tracers only probe surface layers above the bulk mass reservoir\cite{Williams16}. Here we present the first partially spatially resolved observations of the \ccc~$J$=3-2 line emission in the closest protoplanetary disk, TW Hya, a gas tracer that probes the bulk mass distribution. 
Combining it with the \cc~$J$=3-2 emission and the previously detected HD $J$=1-0 flux, we directly constrain the mid-plane temperature and optical depths of gas and dust emission. We report a gas mass distribution of 13$^{+8}_{-5}\times$(R/20.5AU)$^{-0.9^{+0.4}_{-0.3}}$ g cm$^{-2}$ in the expected formation zone of gas and ice giants (5-21\,AU). We find the total gas/millimeter-sized dust mass ratio is 140 in this region, suggesting that at least 2.4\,M$_\oplus$ of dust aggregates have grown to $>$centimeter sizes (and perhaps much larger). The radial distribution of gas mass is consistent with a self-similar viscous disk profile but much flatter than the posterior extrapolation of mass distribution in our own and extrasolar planetary systems.

\end{abstract}


The primary theory for the formation of giant planets is the so-called core accretion scenario, where a rock+ice core forms through the coagulation of planetesimals until it becomes sufficiently massive to accrete a gaseous envelope\cite{Pollack96}. 
In this theory, the spatial distribution of gas in the primitive nebula is not only critical to the later accretion of the atmosphere of giant planets, but also plays an important role in the early planetesimal formation processes as the transport and mixing of grain aggregates depend on the gas turbulence and density\cite{Weidenschilling77b,Birnstiel12}.

\begin{figure}[!h]
\centering
\includegraphics[width=5.5in]{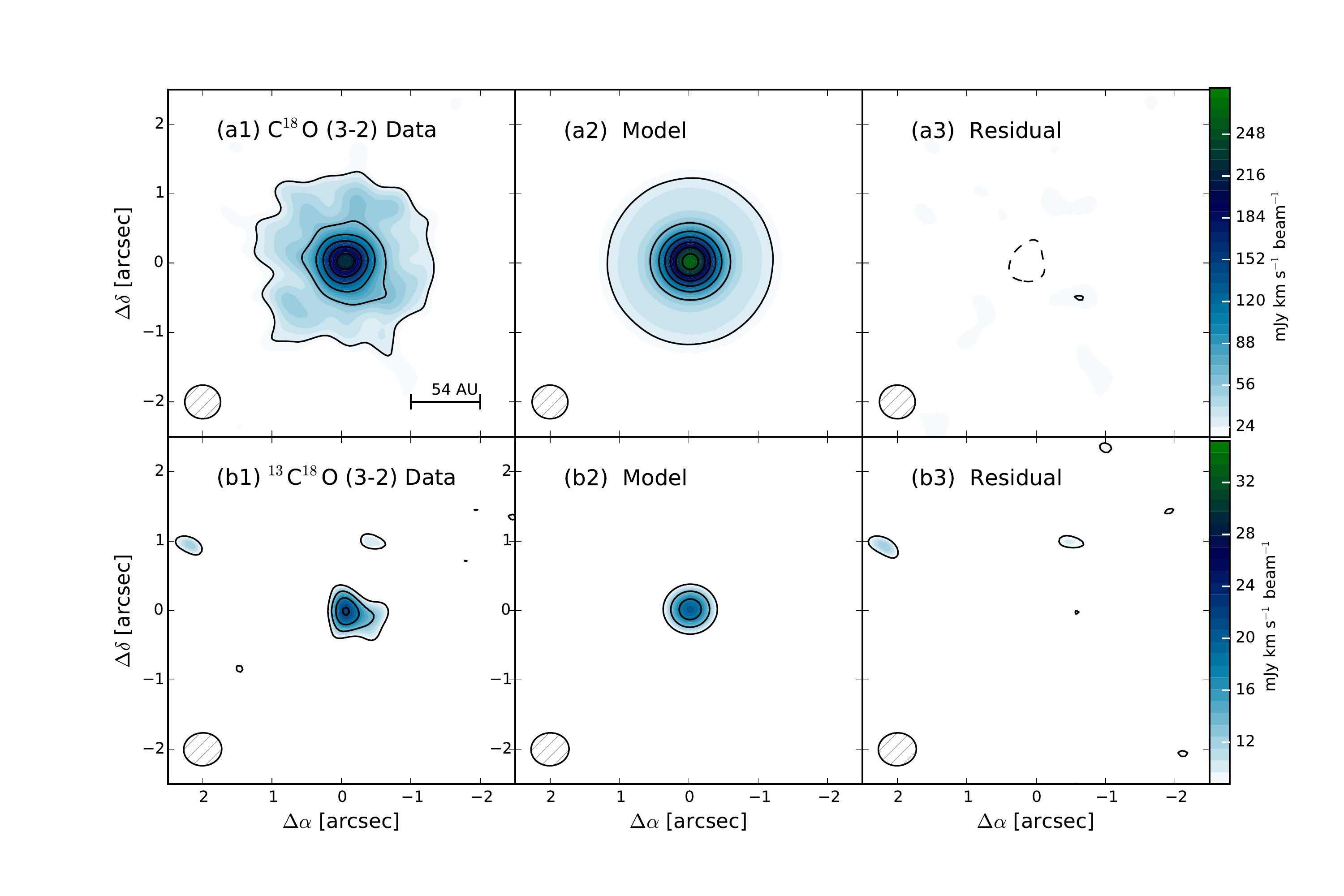}
\vspace{-0.6cm}
\caption{\textbf{ALMA observations of the \cc~and \ccc~(3-2) line emission in the TW Hya protoplanetary disk and the best fitting models.}  (a1, b1)\, Observed Moment zero maps of the \cc~and \ccc~$J$=3-2 line emissions. A bright central peak is seen in both tracers, with a broad diffuse component from the outer disk in the \cc~emission. This morphology is consistent with the theoretical expectation that the mid-plane temperature gradually decreases with radius and that beyond some critical radius most of CO condenses into ice,  where the critical radius is called the CO snowline.  The central peaks of the \ccc~and \cc~emission are 15-40\% larger than the 0.5$^{\prime\prime}$ ($\sim$27.5\,AU) synthesized beam, suggesting that the region inside the CO snowline is partially resolved. In practice, the data provide constraints on smaller spatial scales than the synthesized beam,  since the line emission is also spectrally resolved and the velocity field is known to be Keplerian. 
 The \cc~contour levels start at 3$\sigma$ and increase in 5$\sigma$ steps; the \ccc~contour levels are [2,4,5,6]$\sigma$. (a2,b2) Intensity emission of \cc~and\ccc~(3-2) line emission from the best fitting models. (a3,b3): Imaged residual visibilities by subtracting the best-fitting models from the observations. The dash-lined contours indicate negative values. }
\label{fig:obs}
\vspace{-0.15cm}
\end{figure}

Here we report a high spatial and spectral resolution observational study of the gas mass distribution in the expected formation zone of gas and ice giants using the  Atacama Large Millimeter/submillimeter Array (ALMA). In particular, we robustly detect the \ccc~and \cc~$J$=3-2 line emission (5$\sigma$ and 29$\sigma$, respectively) in the closest  protoplanetary disk around the T Tauri star TW Hya (see Figure~\ref{fig:obs}). TW Hya, at a distance of only$\sim$55\,pc\cite{vanLeeuwen07}, is one of the best studied disk systems and makes a useful analog to the solar nebula: the central star has a stellar mass about 0.6-0.8\,M$_\odot$\cite{Webb99}; the disk itself has a total mass (gas and dust) of $\sim$0.05\,M$_\odot$\cite{Bergin13}, comparable to the total mass of 0.01\,M$_\odot$ in the Minimum Mass Solar Nebula (MMSN) which is the lowest possible mass needed to reproduce the Solar System. The age of TW Hya is 8$\pm$4\,Myr\cite{Donaldson16}, within the expected window for giant planet formation\cite{Chabrier14}.

\begin{figure}[htbp]
\centering
\includegraphics[width=4in]{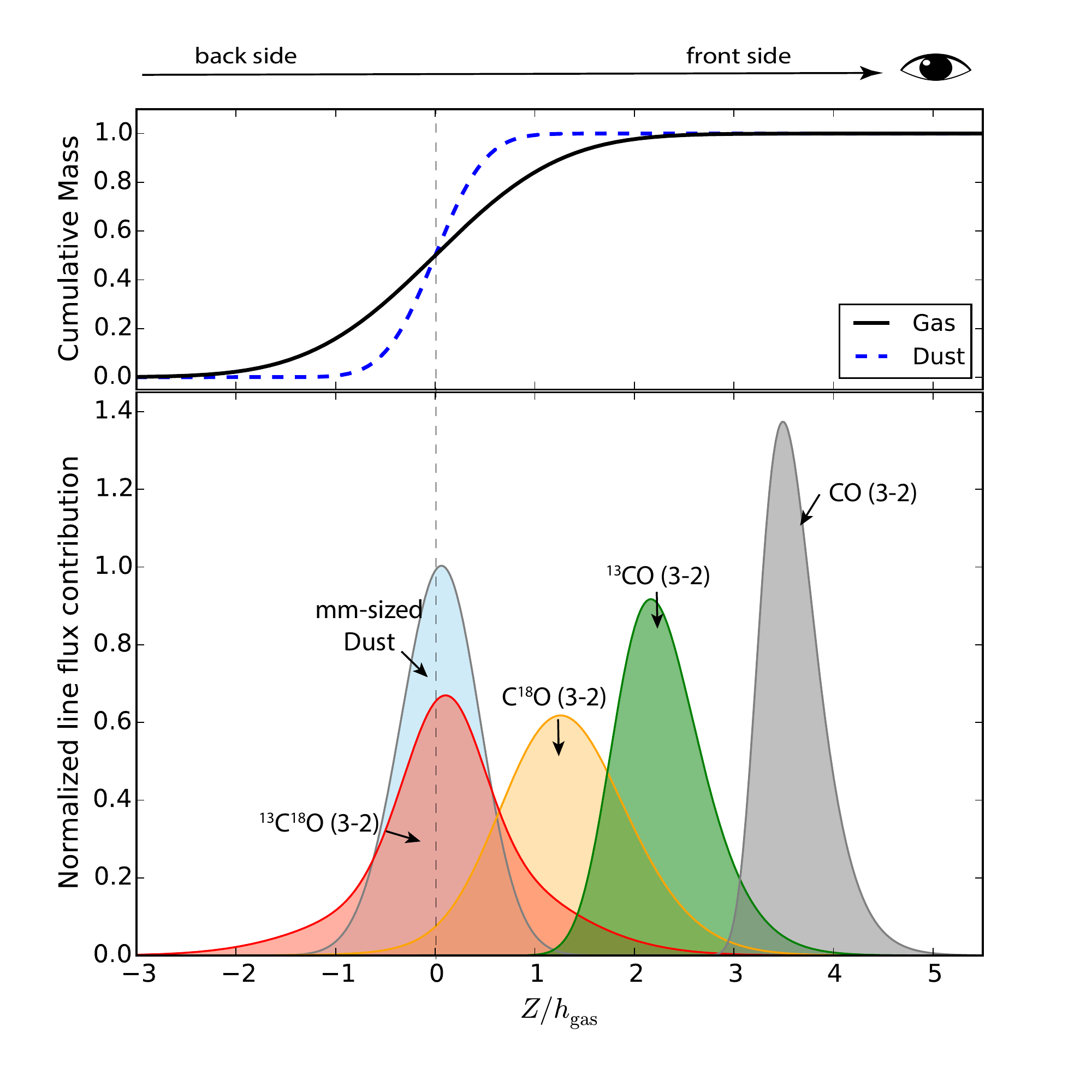}
\vspace{-0.45cm}
\caption{\textbf{Representative vertical contributions of four CO isotopologues in the $J$=3-2 line and 0.93\,mm dust continuum emission in the inner region of the TW Hya disk.} All profiles are normalized to the same total integrated intensity.  Upper panel: Assuming the scale height of the mm-sized dust is 40\% of the gas, the cumulative gas/dust mass distribution from the back side of the disk to the front side. Lower panel: using three typical optical depths for the \ccc, \cc~$J$=3-2 line and continuum emission between 5-20.5\,AU (see Methods), we show that the optically thin \ccc~and dust emission are from both sides of the disk in vertical regions within two gas scale height. Although the optically thick \cc~line emission is mainly from one side of the disk, 40\% its emission is from the vertical region within one scale height and 80\% of its flux are from the region within two scale height. We also show the emitting layers of the highly optically thick $^{13}$CO and CO $J$=3-2 lines 
are from higher regions, which only contain a small fraction of the total surface density.}
\label{fig:cum_flux}
\vspace{-0.35cm}
\end{figure}

These observations were specifically designed to probe the mid-plane interior to 30\,AU. CO,  one of the most volatile species after H$_2$ and He, remains in gas phase even in the mid-plane within a large radius\cite{Qi13}. 
Previous observational studies have used more abundant CO isotopologues (e.g.\,$^{13}$CO), which have stronger line emission,  to study the gas mass distribution in protoplanetary disks\cite{Williams16}. However, inside the mid-plane CO snowline, more abundant CO isotopologue line emission primarily arises from layers at much larger scale heights. As a result, the derived gas mass distribution is highly model dependent, hinging on the adopted temperature and vertical density structures. In Figure~\ref{fig:cum_flux}, we show the vertical emitting regions of CO $J$=3-2 line emission from four isotopologues under representative properties for the TW Hya disk. It shows that only the \ccc~(3-2) line emission arises from both sides of the disk and thus provides a robust probe of the bulk mass.

To accurately constrain physical properties in the mid-plane, we employ a parameterized disk model to reproduce the observations. A full description of the model and the
Markov chain Monte Carlo (MCMC) fitting algorithm is given in the Methods section. The critical assumption of the model is that the \ccc~and \cc~ emission arises predominately from the mid-plane or  immediately adjacent regions, and therefore that the emitting gas can be characterized by a one-dimensional mid-plane temperature. This assumption is reasonable since \cc~and \ccc~are 10$^{-10}$--10$^{-6}$ times less abundant than H$_2$\cite{Wilson99}, and the disk is nearly isothermal within one scale height\cite{Dullemond04}. 
To include potential extinction from dust, we treat the \cc, \ccc~line emission and dust continuum simultaneously.  For the dust emission, we employ the radial profile of surface brightness derived by Andrews et al. (2016)\cite{Andrews16}, because these continuum observations were taken at a similar frequency as our continuum data (345 v.s.\,320\,GHz) but with much higher spatial resolution ($\sim$1\,AU). 
We use simple power-law functions to characterize the radial profiles of the temperature and integrated line strength. 
The best-fitting parameters and their uncertainty are listed in Table~\ref{tab:best_para}, and the best-fitting model is plotted in Figure~\ref{fig:obs} in comparison with the data.

\begin{table}[!t]
{\footnotesize
\caption{Best fitting TW Hya disk model parameters}

\begin{center}
\begin{tabular}{ccccccccccc}
\hline
\hline
R$_{\rm snow}$ & T$_{\rm snow}$ & $q_t$& $\tau_0$ & $q_1$ & $q_2$ & log$\delta$ &  $^{12}$C/$^{13}$C &R$_{\rm ring}$&A$_{\rm ring}$ & $\sigma_{\rm ring}$ \\

(AU) & (K) & & (km s$^{-1}$)& & & &  &(AU) & &(AU) \\
\hline
20.5$\pm1.3$ & 27$^{+3}_{-2}$ &-0.47$^{+0.06}_{-0.07}$ & 2.5$^{+0.6}_{-0.5}$ & -0.18$^{+0.39}_{-0.23}$ &
-1.9$\pm0.1$ &-0.94$\pm0.14$ & 40$^{+9}_{-6}$ & 60$^{+1}_{-1}$&
6.3$^{+1.2}_{-0.8}$ &11$\pm1$\\
 \hline
\end{tabular}
\end{center}

{\rm Note: R$_{\rm snow}$--the mid-plane CO snowline location. T$_{\rm snow}$--the mid-plane temperature at R$_{\rm snow}$.  $q_t$--the power index that characterizes the mid-plane temperature for radii inside the CO snowline, T(r) = T$_{\rm snow} \times(r/r_{\rm snow})^{q_t}$. $\tau_0$--integrated line strength at R$_{\rm snow}$,  $q_1$ and $q_2$--the power indices that characterize how the integrated line intensity varies with radius inside R$_{\rm snow}$ and outside R$_{\rm snow}$, respectively. log$\delta$: $\delta$ is the depletion factor of the integrated line strength across the R$_{\rm snow}$. $^{12}$C/$^{13}$C: isotopologue ratio. \{R$_{\rm ring}$, A$_{\rm ring}$, $\sigma_{\rm ring}$\}--three parameters that characterize the ring structure seen in the \cc~emission.} 

\label{tab:best_para}
}
\end{table}

\begin{figure}[p!]
\centering
\includegraphics[width=5in]{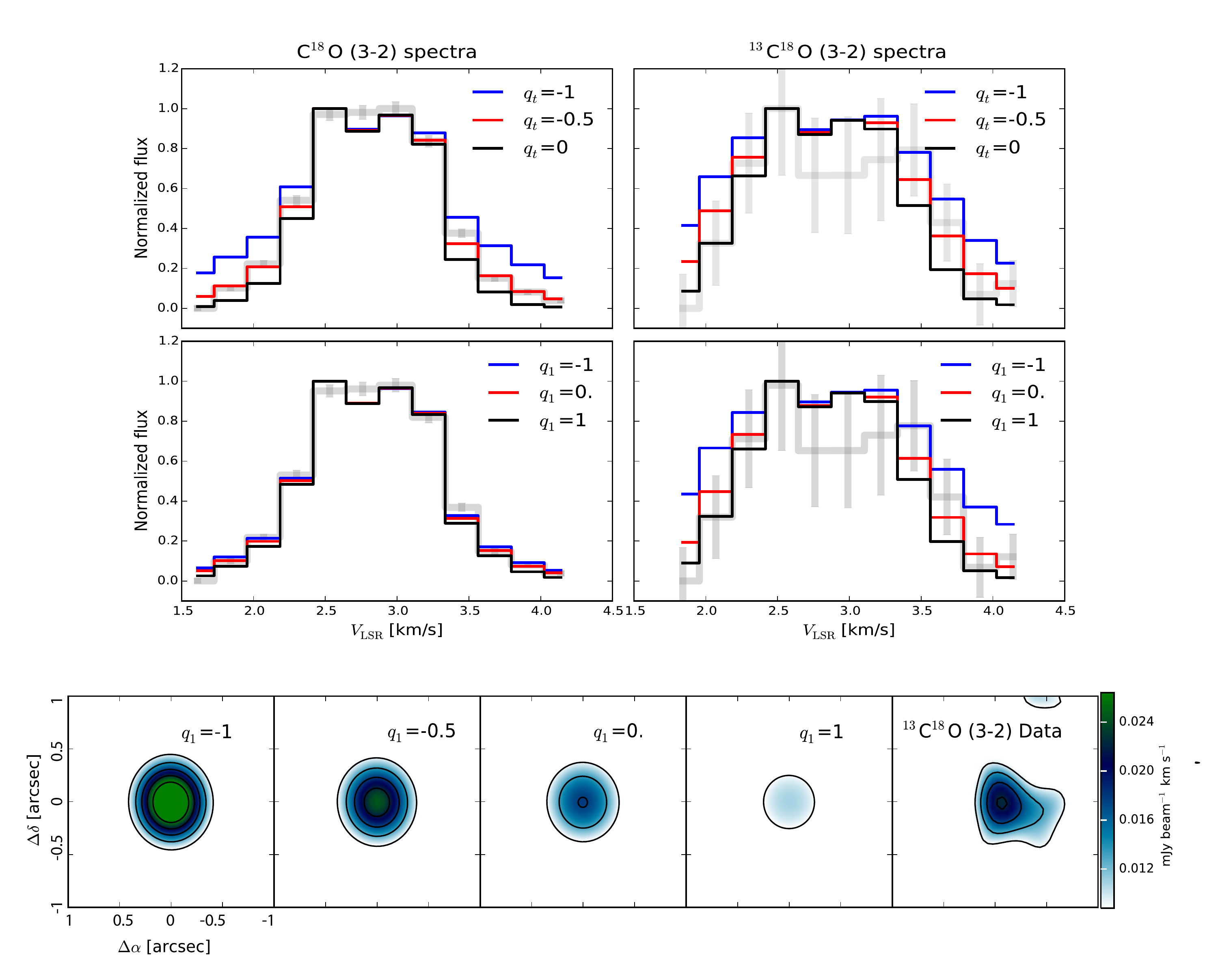}
\vspace{-0.45cm}
\caption{Comparison of models with different temperature and optical depth distributions with the \ccc~and \cc~(3-2) line spectra and  \ccc~(3-2) moment zero map. All models adopt the best-fitting model parameters listed in Table 1 except for the q$_t$ or q$_1$ values. The grey lines in the upper panel plots show line spectra and their uncertainty from the observations. In all plots, the flux at the central channel has been normalized to match with the best-fitting model. It can be seen that the spectral of the \cc~is sensitive to temperature distribution ($q_t$) but not the optical depth distribution ($q_1$), and \ccc~(3-2) spectra and moment zero map is sensitive to both the temperature and optical depth distribution. }
\label{fig:sensitivity}
\vspace{-0.15cm}
\end{figure}

\begin{figure}[p!]
\centering
\includegraphics[width=3.7in]{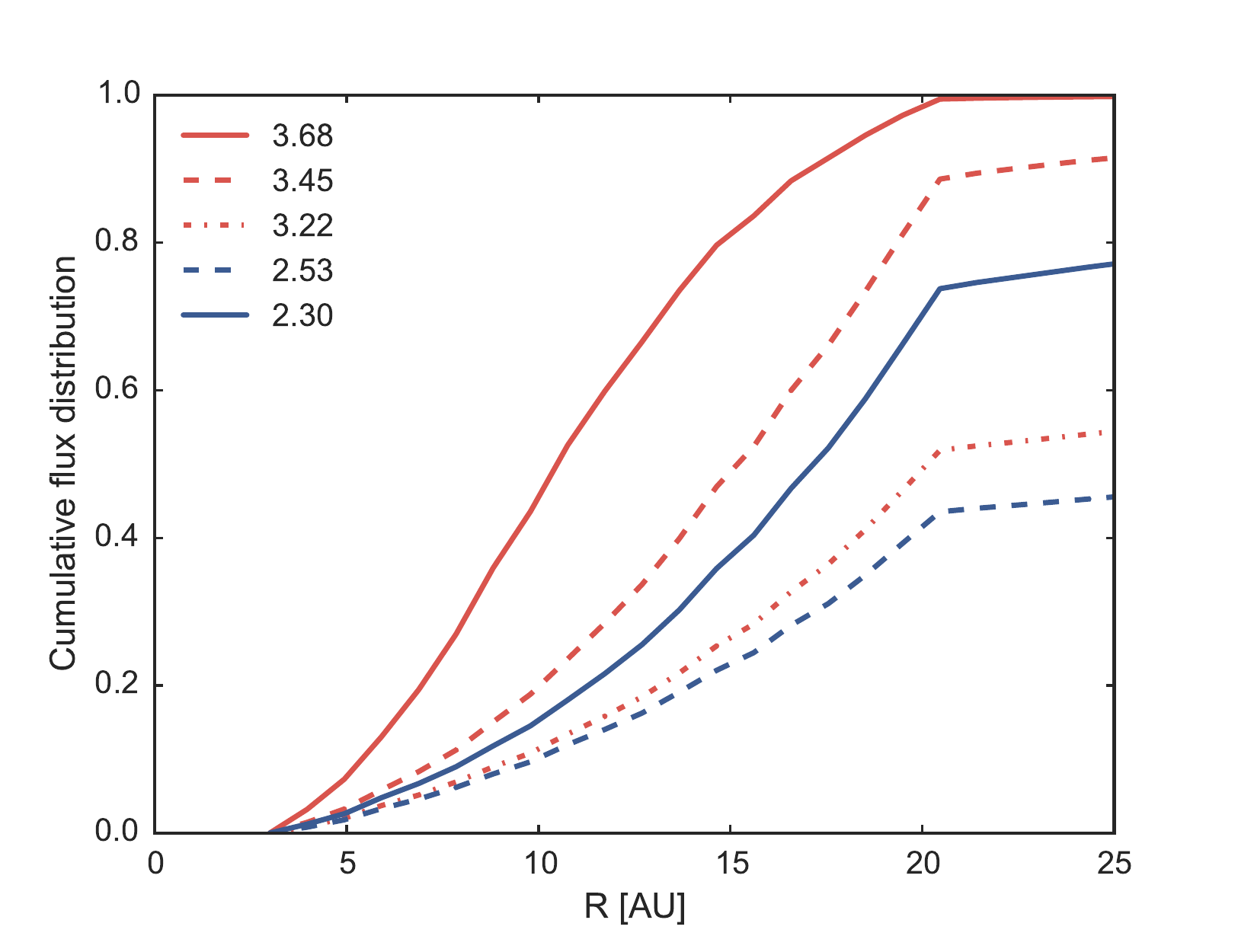}
\vspace{-0.45cm}
\caption{Cumulative flux distributions of the \ccc\,(3-2) line emission in the five velocity channels that have 3\,$\sigma$ detections, based on the best-fitting model. The LSR velocities of these channels are labelled at the upper left corner. Due to the regular Keplerian rotation velocity field in the disk, the radial dependence of flux contribution in a velocity channel varies from channel to channel. The figure shows that the 40-100\% of the line emissions in these channels arise from regions inside the 21\,AU, and indicates that each channel samples emissions at different radii with different weights. Therefore the radial distribution of the emission inside 21\,AU can be constrained even the emission is only partially spatially resolved. The inner limit is $\sim$5\,AU based on current observations (10\% flux contribution in the 3.68\,km s$^{-1}$ channel).  
}
\label{fig:f_cum}
\vspace{-0.15cm}
\end{figure}

We report the first direct measurement of the mid-plane CO snowline at 20.5$\pm1.3$\,AU, in the TW Hya disk. The \ccc~$J$=3-2 emission is optically thin ($\tau<$0.2) throughout the disk, and the dust continuum emission has $\tau\sim$0.5, suggesting that the \ccc~line emission arises from both sides of the disk and thus provides model-independent constraints on the mass distribution. The \cc~emission is optically thick ($\tau\sim$8) inside 20.5\,AU and therefore provides direct constraints on the gas temperature rather than the mass distribution (Supplementary Figure~\ref{fig:sensitivity}). The observations trace the mid-plane between 5-20.5\,AU: the outer limit is the mid-plane CO snowline location (beyond that CO condenses at the mid-plane); the inner boundary is limited by the radius of 10\% cumulative flux at the highest velocity channel with a \ccc~ emission detection (Methods and Supplementary Figure~\ref{fig:f_cum}).

\begin{figure}[!ht]
\centering
\includegraphics[width=5in]{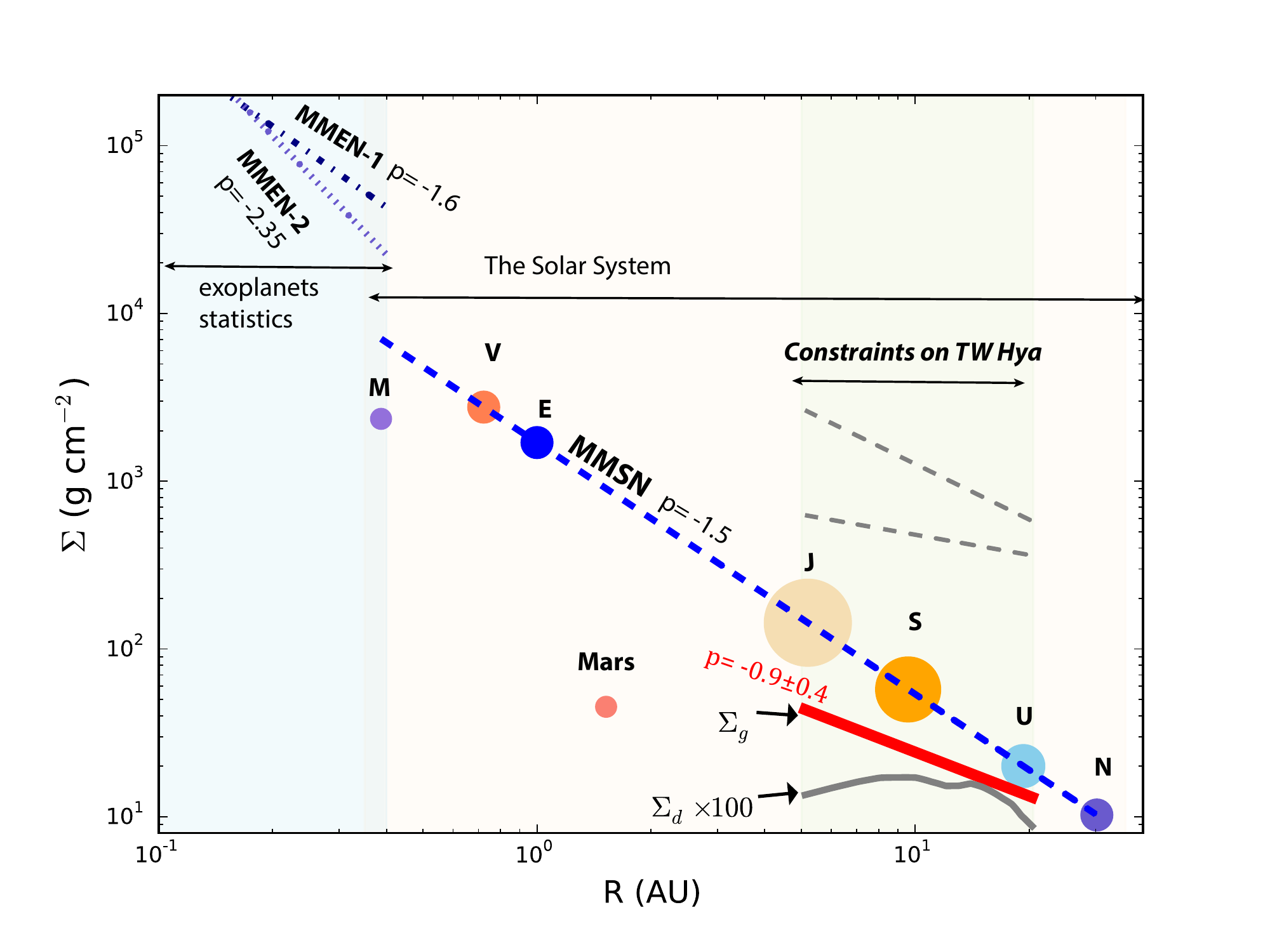}
 \vspace{-1cm}
\caption{\textbf{Comparisons of the gas mass distribution in the TW Hya disk with models of in situ planet formation (MMSN/MMEN).} The traditional Minimum Mass Solar Nebula (MMSN) model is derived by spreading planetary mass into a series of annuluses centered on the current planet locations and then adding the light element mass needed to restore a solar composition\cite{Weidenschilling77, Hayashi81}. Here we show the MMSN model of Hayashi et al.\,(1981)\cite{Hayashi81}, $\Sigma_g = 1700\times(r/1AU)^{-1.5}$.  The analog of MMSN in extrasolar systems is called the Minimum Mass Extrasolar Nebula (MMEN), based on the statistics of the close-in super-Earths population (Period$<$100\,days).  Two versions of MMEN have been derived by assuming different feeding zone widths: Chiang \& Laughlin (2013)\cite{Chiang13} (MMEN-1, p=-1.6) and Schlichting et al.\,(2014)\cite{Schlichting14} (MMEN-2, p= -2.35). Here we show that the gas surface density profile in the TW Hya disk over 5-20.5\,AU (red solid line, p =-0.9$^{+0.4}_{-0.3}$) is flatter than all of the MMSN/MMEN models. The two gray dashed lines indicate the $\pm$1$\sigma$ range of the power index (offset from the best-fit model for clarity). In the same region, the surface density of millimeter-sized dust grains (gray solid line) is $\sim$140 times less than the gas surface density. To reconcile these data with the canonic gas/dust mass ratio of 100 in the ISM, at least 2.4\,M$_{\oplus}$ of solids need to have grown into cm or larger sized bodies.}
\label{fig:mmsn}
\vspace{-0.25cm}
\end{figure}

With the temperature and optical depths from the best-fitting model,  we can derive the column density distribution of \ccc~and then the gas mass distribution for a given \ccc~abundance~($x_{\rm ^{13}C^{18}O}$). Here we employ the spatially unresolved HD $J$=1-0 line flux of the TW Hya disk to estimate the $x_{\rm ^{13}C^{18}O}$\cite{Bergin13}.  Combining the uncertainties in our model parameters and in the HD line flux, we find $x_{\rm ^{13}C^{18}O}$=1.7$^{+1.3}_{-0.8}\times$10$^{-10}$ and  $\Sigma_{\rm gas}$=13$^{+8}_{-5}\times$(R/20.5\,AU)$^{-0.87^{+0.38}_{-0.26}}$\,g cm$^{-2}$ between 5-20.5\,AU (for details, see Methods). 

\begin{figure}[p!]
\centering
\includegraphics[width=7in]{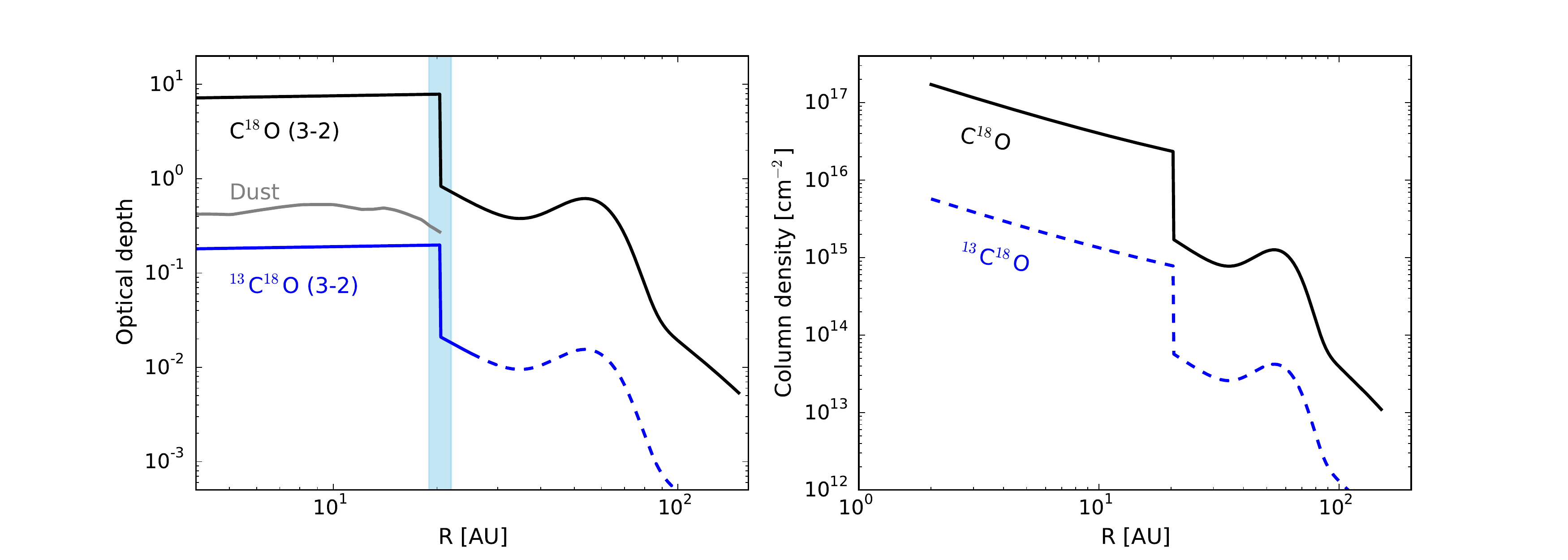}
\vspace{-1cm}
\caption{Left: Optical depth at the center of the local line profile for the \cc~(black) and \ccc~(blue) $J$=3-2 transitions, and for the 0.93\,mm continuum emission. The optical depth of dust is only considered for the region within the CO snowline, because beyond that the CO line emission is mainly from a warm layer higher than the mid-plane. Right: Column density distribution of \cc~and \ccc.}
\label{fig:tau}
\vspace{-0.35cm}
\end{figure}

For dust mass, the conversion from optical depth to surface density depends on the dust opacity $\kappa_{\nu}$, which may vary with radius due to different levels of dust growth. 
The growth level of dust aggregation is usually estimated with a parameter $\beta$ where $\kappa_{\nu}\propto\nu^{\beta}$. Combining our best-fitting  temperature structure and previous continuum observations at 145, 233, and 345\,GHz\cite{Andrews16, Tsukagoshi16}, we find that the $\beta_{\rm 145-233\,GHz }$ increases from $\sim$ 0 to 1.1 between 5-20.5\,AU, consistent with the finding of Tsukagoshi et al. (2016) which used a different temperature structure. But we also find that the $\beta_{\rm 233-345\,GHz }$ is roughly flat in the same region.  As the dust emission in this region has $\tau<1$ (Supplementary Figure~\ref{fig:tau}), the $\beta\lesssim1$ suggests that significant fraction of dust has grown to mm-size or larger\cite{Draine06}, and the discrepancy of $\beta$ among different wavelengths hints that the dust size distribution in this region is possibly significantly different from the typically assumed power-law function with an index of -3.5. Because detailed modeling of the dust size distribution is beyond the scope of this paper, here we adopt a constant $\kappa$(0.93\,mm) = 3.1\,cm$^2$g$^{-1}$ to estimate the dust mass of grains no larger than millimeter.  The opacities adopted are similar to those used in previous models. Assuming a constant $\kappa_\nu$, the radial distribution of mm-sized dust mass between 5-20.5\,AU is shown in Figure\,\ref{fig:mmsn}. 

We measure the total gas-to-mm-sized dust ratio is 140$^{+43}_{-31}$ for the 5-20.5\,AU region, much higher than the typical ratio of 100 in molecular clouds. To recover a gas-to-dust ratio of 100, we find that 2.4$^{+3.2}_{-2.0}$\,M$_{\oplus}$ solid masses are missing. This estimation is actually a lower limit, because the dust radial drift to the inner region should in general decrease the total gas-to-dust ratio in the inner disk\cite{Birnstiel12} and because the overall gas surface density also decreases with time. Therefore more than 2.4\,M$_{\oplus}$ masses of solids may have grown to at least centimeter-sized and perhaps much larger bodies in the TW Hya disk. This estimation sheds new light on the previous skepticism of planet formation in the TW Hya disk.  Recently, a series of dark rings were discovered in the (sub)mm continuum emission of the TW Hya\cite{Andrews16} and other disks\cite{alma15}, which has excited heated debates on the origin of these mysterious rings and especially whether they formed via planet-disk interaction\cite{Zhang16, Dipierro15}. Simulations suggest that the rings are too shallow and narrow to be gaps carved by Jupiter-like planets, but smaller bodies are still possible\cite{Jin16}.

The direct measurement of gas mass distribution in the TW Hya disk provides the first data point to compare with the solar nebula over a similar radial region. 
Compared to the traditional MMSN, the TW Hya disk has less mass between 5-21\,AU, though the central star is slightly less massive than the Sun. More importantly, the mass distribution in TW Hya is much flatter in this region (Figure~\ref{fig:mmsn}). Extending the comparison to exoplanetary systems, the MMSN has its counterpart the MMEN, which is based on statistics of the close-in super-Earth population ($<$0.4\,AU). The mass distribution in the MMEN models is at least as steep as the MMSN and perhaps much steeper (from -1.6 to -2.35), depending on the assumption of feeding zone size\cite{Chiang13,Schlichting14}. 
 
Although more observations are needed to determine whether the gas mass distribution in TW Hya is common in protoplanetary disks, its shallow nature follows that derived from the dust emission\cite{Andrews07,Isella09}, and the apparent difference  questions the use of MMSN/MMEN models as a standard initial mass distribution. The fundamental assumptions of these models are that planets formed at their current location and that the metal-to-(H+He) elemental ratio is solar throughout the solar nebula. In fact, the in situ formation assumption has been challenged by the discovery of hot Jupiters in extrasolar systems, where these close-orbit giant planets are too massive to form at their current location and probably are results of inward migration\cite{Wright12}. However, Solar system formation models that include planet migration require even steeper initial mass distribution than the MMSN\cite{Desch07}. 
If the gas mass distribution of the solar nebula was similar to the TW Hya disk,  this large discrepancy means that the mass distribution of planetesimals (represented by planetary mass distribution before migration) must entirely decouple from the gas mass distribution in the solar nebula. In short, the MMSN/MMEN models may offer useful estimation as to the total mass of the natal protoplanetary disks but should not be used as quantitative templates of initial mass distribution for planet formation models. 

Ultimately, the quantitive prediction of the mass distribution in protoplanetary disks should come from a detailed understanding of angular momentum transport in disks, for which observational studies such as this provide useful constraints. For example, in the self-similar solution of accretion disks\cite{Pringle81}, the power-law index $\gamma$ of the radial variation of viscosity (where $\nu\propto r^\gamma$) equals the power-law index ($p$) of surface density ($\Sigma_{\rm gas}\propto r^{-p}$ in the region r$\ll$ characteristic radius of the disk ).  If the disk viscosity is mainly driven by turbulence (presumably due to magnetorotational instability), the effective kinematic viscosity is usually characterized by the Shakura-Sunyaev $\alpha$-model: $\nu=\alpha c_s^2/\Omega$, where $c_s$ is the isothermal sound speed ($c_s^2\propto T_{\rm gas}$) and $\Omega$ is the Keplerian angular velocity. Taking our measurement of $T_{\rm gas}$ and $\Sigma_{\rm gas}$, we find an $\alpha\propto r^{0.16\pm0.39}$ between 5-21\,AU. This result is consistent with the typically used constant $\alpha$ models, within uncertainties. 

With the TW Hya measurement alone, however, it is not yet possible to test realistic magnetohydrodynamic models of disk evolution, because current simulations predict that the surface density profile varies widely depending on the input parameters and age of the disk\cite{Bai16, Suzuki16}. In the next few years, these models will be tested by measuring mass surface density profiles in a sample of disks with different ages and masses using techniques such as those describe here.


\newcommand{\nat}{{ Nature }}
\newcommand{\aap}{{Astron. \& Astrophys. }}
\newcommand{\aj}{{ Astron.~J. }}
\newcommand{\apj}{{ Astrophys.~J. }}
\newcommand{\araa}{{Ann. Rev. Astron. Astrophys. }}
\newcommand{\apjl}{{Astrophys.~J.~Letters }}
\newcommand{\apjs}{{Astrophys.~J.~Suppl. }}
\newcommand{\apss}{{Astrophys.~Space~Sci. }}
\newcommand{\icarus}{{Icarus }}
\newcommand{\mnras}{{MNRAS }}
\newcommand{\pasp}{{ Pub. Astron. Soc. Pacific }}
\newcommand{\pasj}{{ Pub. Astron. Soc. Japan }}
\newcommand{\ssr}{{Space Sci. Rev.}}
\newcommand{\planss}{{Plan. Space Sci. }}
\newcommand{\physrep}{{ Phys. Rep.}}
\newcommand{\bain}{{Bull.~Astron.~Inst.~Netherlands }}
\newcommand{\pra}{{Phys.~Rev.~A}}%



\begin{addendum}
 \item This
paper makes use of the following ALMA data: ADS/JAO.ALMA\#2015.1.00308.S. 
ALMA is a partnership of ESO (representing its member states), NSF (USA) and NINS 
(Japan), together with NRC (Canada), NSC and ASIAA (Taiwan), and KASI (Republic of Korea),
 in cooperation with the Republic of Chile. The Joint ALMA Observatory is operated by ESO, AUI/NRAO and NAOJ. 
 We thank Takashi Tsukagoshi for sharing radial brightness temperature profiles of ALMA 145 and 233\,GHz continuum observations of the TW Hya disk. 
This work was supported by funding from the National Science Foundation grant AST-1514670 and NASA NNX16AB48G. 
L.I.C. acknowledges the support of NASA through Hubble Fellowship grant HST-HF2-51356.001.
 
 \item[Author Contributions]  K.Z. led the data processing, analysis, and manuscript preparation. E.A.B. led the preparation of the observing proposal and  K.R.S. assisted with the parameterized modeling. All authors
were participants in the elaborating the observing proposal, discussion of results, determination of the
conclusions, and revision of the manuscript.

 \item[Competing Interests] The authors declare that they have no
competing financial interests.
 \item[Correspondence] Correspondence and requests for materials
should be addressed to Ke Zhang~(email: kezhang@umich.edu).
\end{addendum}

 
\begin{methods}

\subsection{ALMA observations.}

TW Hya was observed with the Atacama Large Millimeter/submillimeter Array (ALMA) as a part of the Cycle 3 proposal 2015.1.00308.S (P.I. Bergin). Observations were carried out on 2015 March 8, with 40 antennas and baselines of 15-460\,m (16-492 \kl). The total on-source integration time was 67.5 min. The nearby quasar J1058+0133 was used for bandpass calibration, and J1037-2934 for both flux and phase calibration. The correlator was 
configured to simultaneously observe four spectral windows (SPWs). Three of the SPWs were used for line observations with a resolution of $\delta\nu$ = 122\,kHz (a spectral resolution of 0.23\,\kms~after the default Hanning smoothing is applied), and one for dedicated continuum observations. The \ccc~$J$=3-2 line was placed in SPW 1, and \cc~3-2 line in SPW 2 and 3. The three line SPWs are all 117.2\,MHz wide and start at 314.171, 329.319 and 329.264\,GHz, respectively.  The SPW 0 was used for continuum observations with a total bandwidth of 1.875\,GHz. 

The visibility data were calibrated by ALMA staff following standard procedures with the pipeline. We further performed self-calibration on the calibrated data in phase and amplitude in CASA 4.6.12, using the dedicated continuum observations in SPW 0. The self-calibrated phase and gain solutions were then applied to the three line SPWs. The data were subsequently continuum subtracted using line free channels. We generated synthesized images using  a CLEAN deconvolution with a manual mask (Briggs weighting with a robust parameter of 0.5). For the 321\,GHz continuum emission map, the synthesized beam is 0.54$^{\prime\prime}\times0.47^{\prime\prime}$, with a position angle of -87 degrees. The \cc~line image cube has a noise level of 6.5\,mJy beam$^{-1}$ in each 0.23\,\kms~wide velocity bin, and the \ccc~line image cube has a noise level of 4\,mJy beam
$^{-1}$. The peak signal-to-noise ratio in the moment zero maps is 5 for \ccc~ and 29 for \cc, respectively. The synthesized images of line and continuum emission are displayed in Figure~\ref{fig:obs}.

The observations of \ccc~and \cc~were designed to probe the mid-plane interior to 30\,AU. In the radial direction, CO remains in gas phase even in the mid-plane within a large radius\cite{Qi13}. In the vertical direction, previous studies have shown that the most abundant CO isotopologues ($^{12}$CO and $^{13}$CO) readily become optically thick near the surface layer of protoplanetary disks and thus do not provide constraints on the deep interior \cite{Beckwith93, Pietu07}. The two CO isotopologues chosen, \ccc~and \cc, are on average 38,433 and 577 times less abundant than $^{12}$CO in the interstellar medium (ISM)\cite{Wilson99}. Their optical depths are correspondingly much smaller than the $^{12}$CO and $^{13}$CO lines, and thus can probe much deeper regions in protoplanetary disks.

\subsection{Parameterized disk model.}


Our goal is to constrain three key physical properties of the inner region of the TW Hya disk: the location of the mid-plane CO snowline (r$_{\rm snow}$), the mid-plane temperature structure inside r$_{\rm snow}$,  and the radial mass distribution of the gas.  To constrain these physical properties, we use an axisymmetric 1-Dimensional model to reproduce the \ccc~and \cc~$J$=3-2 line emissions and dust continuum emissions at 345\,GHz. The axisymmetric approximation is valid since no clearly asymmetric features have been seen in the line and continuum emissions from the TW Hya disk to date. The model structure consists of two distinct radial zones: inside and outside the mid-plane CO snowline (see Supplementary Figure~\ref{fig:tau_model}). The two zones are used to model the break of the CO gas column density distribution due to condensation. 

Our key focus is the nature of the region inside r$_{\rm snow}$.  The r$_{\rm snow}$ location is measured from the size of the bright central peaks in the \ccc~and \cc~ images.  We assume that inside r$_{\rm snow}$, the \cc, \ccc~3-2 line and continuum emissions primarily arise from the mid-plane or immediately adjoint regions, and therefore that the emitting gas column at each radius can be characterized by the mid-plane temperature. We will discuss the self-consistency of this assumption for our best-fitting model later.  The abundance ratio of \cc/\ccc~is assumed to be constant throughout the disk and modeled as a free parameter\cite{Miotello14}. 

Under these two assumptions, the temperature and optical depths of line/continuum emission can be independently constrained by the \ccc~and \cc~$J$=3-2 line and continuum emissions. The observed \cc~line emission at the image center is only eight times stronger than the \ccc~line emission, much smaller than the typical ISM ratio of $\sim$69. The \cc~line emission is thus likely to be optically thick, which provides direct constraints on the temperature inside the r$_{\rm snow}$ (see Supplementary Figure~\ref{fig:sensitivity}).  With the temperature structure constrained, the optically thin \ccc~line emissions provide constraints on the CO gas column density distribution, and the continuum emissions constrains the dust optical depth. The CO gas column density distribution derived is then used as a proxy of the gas mass distribution, as current chemical models suggest that the CO/H$_2$ abundance ratio is roughly constant in disk regions warmer than the CO condensation temperature\cite{Walsh10}. The absolute scale factor is constrained by the previously observed HD (1-0) line flux\cite{Bergin13}.

The detailed construction of the model is described below. Inside r$_{\rm snow}$, the continuum-subtracted line intensity and continuum intensity at each radius are modeled as an isothermal slab with the following equations: 
\begin{equation}
\left\{
\begin{aligned}
I_d     &=B_\nu(T) (1-e^{-\tau_d})\\
I_{\rm C^{18}O} &=B_\nu(T) (1-e^{-(\tau_d+\tau_{C^{18}O})})-I_d\\
I_{\rm ^{13}C^{18}O} &= B_\nu(T) (1-e^{-(\tau_d+\tau_{^{13}C^{18}O})})-I_d\\
                            &= B_\nu(T) (1-e^{-(\tau_d+\tau_{C^{18}O}/ratio)})-I_d
                            \end{aligned}
\right.                            
\end{equation}
where $\tau_{\rm C^{18}O}(r,v)   = \tau_0\times (r/r_{\rm snow})^{q_1}\times\phi_v$ (see Supplementary Figure~\ref{fig:tau_model}), and $\phi_v$ is the local line profile that is modeled as a Gaussian with a standard deviation \ $\sigma_{\rm V} = \sqrt{2k_{\rm B}T/m_{\rm CO} +v_{\rm turb}^2}$. We fix  $v_{\rm turb}$  at 0.01\,km s$^{-1}$, a turbulence level consistent with an $\alpha$ of 10$^{-3}$ at 30\,K, as previous ionization studies of the TW Hya disk\cite{Cleeves15} suggest that the turbulence level in the inner 50\,AU region is likely to be very low. In addition, we test several models with different values of $v_{\rm turb}$. We find that using smaller values of $v_{\rm turb}$ does not change our best fitting results.  Using a much larger $v_{\rm turb}$\cite{Teague16}  (0.1\,km s$^{-1}$) would make the surface density profile slightly shallower but the difference is three times smaller than our uncertainty. We model the macroscopic velocity field as a Keplerian velocity field\cite{Flaherty15}, and adopt stellar properties and disk inclination/position angle from the literature. The fixed model parameters are listed in Supplementary Table~\ref{tab:sup1}. 

Outside r$_{\rm snow}$, the CO line emissions arise from a warm layer above the mid-plane dust emission and thus the temperature of the dust continuum and the two CO isotopologues are significantly different. As previous studies have shown that the (sub)mm continuum emission is optically thin\cite{Hogerheijde16}, we ignore dust emission in the outer disk. We use a constant temperature of 21\,K for the CO isotopologues emission, as previously measured from CO isotopologue line emissions in the outer disk region of the TW Hya disk\cite{Schwarz16}.  For the line optical depth, we use $\tau_{\rm C^{18}O}(r,v)   = \delta\times\tau_0\times(r/r_{\rm snow})^{q_2}\times\left(1+Ae^{-\frac{(r-r_{\rm ring})^2}{2\sigma_{\rm ring}^2}}\right)\times e^{-r/100AU}\times\phi_v$, where a Gaussian ring  is added to model the ring structure around 60\,AU seen in the \cc~(3-2) line emission\cite{Nomura16,Schwarz16}, and the  exponential component ($e^{-r/100AU}$) is added to model the apparently sharp truncation of the CO emission beyond 100\,AU.

For the continuum emission, we employ the radial profile of surface brightness derived from the high spatial resolution dust continuum observations by Andrews et al. (2016)\cite{Andrews16}, which were taken at a similar frequency (345\,GHz) and with a much higher spatial resolution ($\sim$1\,AU). We note that the spatial resolution of 
our constraints on the dust optical depth profile is 5-10\,AU, because it is limited by the resolution of the temperature distribution measured from the \cc~line emission (see Supplementary Figure~\ref{fig:sensitivity}).

To compare models with observations, we carry out Fourier transforms on the model image cubes and retrieve the complex visibilities at the ($u,v$)-plane locations sampled by the ALMA observations  using the Python package \textsc{vis\_sample}\footnote{Available at \url{https://github.com/AstroChem/vis_sample}}. The modeled visibilities can then be directly compared with observed visibilities from ALMA observations. To sample the posterior probability distribution for each parameter, we use an MCMC approach. In practice, the sampling exploration is carried out by the affine-invariant routine EMCEE\cite{Foreman13}. The prior distribution of parameters is assumed to be uniform across a wide parameter space. Our chains consist of 100 walkers and 1000 steps. 
The best-fit parameters and their uncertainties are listed in Table~\ref{tab:best_para} and direct comparison of the observed \cc~and \ccc~$J$=3-2 channel maps with synthetic data from the best model are plotted in Supplementary Figure~\ref{fig:cc} and \ref{fig:ccc}.

\subsection{CO snowline or jump in gas surface density.} To investigate whether the jump observed in the CO column density can be caused by a jump in the gas column density rather than the CO mid-plane snowline, we include two additional models of the gas mass distribution (Model 2 and 3). The two additional models are based on the assumption the gas-to-dust mass ratio and the dust opacity are constant throughout the disk. Model 2 is the gas surface density distribution derived by van boekel et al. (2016)\cite{vanBoekel16} that fits the scattered light images of the TW Hya disk at 0.63, 0.79, 1.24 and 1.62\,$\mu$m (expected to trace micron-sized dust grains that are coupled with the gas). Model 3 is based on the observed surface brightness distribution of the 345\,GHz continuum emission of the TW Hya disk. Using the mid-plane temperature profile we measured from the \cc\,(2-1) line emission, we derive a surface density of the mm-sized dust particles and scale it by 100 to represent the gas mass distribution. The two models as well as our best-fitting model for the CO lines (Model 1) are plotted in the Supplementary Figure~\ref{fig:f_model}. 
For the three models, we show the radial surface brightness profiles of the \cc\, (2-1) moment zero map and the line emissions at the 3.22 km/s velocity channel.  We find that neither models based on dust observations (Model 2 and 3) could reproduce the observed \cc~emissions.   
Therefore the CO column density jump cannot be explained by a jump in the gas surface density profile and thus chemical effect must be included. Considering that our best-fitting model requires an order of magnitude decreases in the CO column density and the temperature measured is consistent with the condensation temperature of CO onto water ice surface, the jump in CO gas column density is most likely caused by a mid-plane CO snowline.  

\subsection{Comparison with previous models.} 
The retrieved physical properties at the mid-plane show some differences compared to previous model predictions.  We measured a mid-plane CO snowline at 20.5$\pm1.3$\,AU from the spatial resolved \cc~and \ccc~(3-2) images, which is closer to the central star than the previous estimation of 30\,AU that derived from the N$_2$H$^{+}$ (4-3) line emission as an indirect tracer\cite{Qi13}. This discrepancy is not surprising as Van't hoff et al. (2016) showed that N$_2$H$^{+}$ line emissions provide only an upper limit to the mid-plane CO snowline radius\cite{vantHoff16}. Moreover, our CO snowline location is consistent with the prediction of 19\,AU from the Van't hoff et al. (2016) chemical models of the TW Hya disk. The mid-plane temperature at the CO snowline is  27$^{+4}_{-3}$\,K, consistent with the expected freeze-out temperature when CO condenses onto a surface formed by a mixture of water and CO$_2$ ice\cite{Cleeves14}. 

We find that the dust emission has $\tau\sim$0.5 inside 20\,AU, about a factor of 2-4 less than previous global models for the TW Hya disk\cite{Hogerheijde16}.  The previous models, however, were based on continuum observations that did not resolve the inner 20\,AU (beam size $\sim$20-25\,AU) and a temperature profile derived from spectral energy distribution rather than direct measurements, which makes the estimation for the inner 20\,AU zone highly uncertain. Moreover, if the dust emission is already optically thick, its brightness temperature at the CO snowline would be equal to the CO condensation temperature. However, Andrews et al.\,(2016)\cite{Andrews16} showed that the brightness temperature is only 12\,K at 20\,AU, much cooler than realistic condensation temperature (21-46\,K) in a high-pressure environment such as the disk mid-plane. Therefore, the dust emission inside 20\,AU is possibly has smaller optical depth than previously assumed.

We measure a $^{12}$C/$^{13}$C ratio of 40$^{+9}_{-6}$ based on the optical depth ratio of the \cc/\ccc~(3-2) line emissions. This ratio is smaller than the average ratio in local ISM of 69$\pm$6\cite{Wilson99} but consistent with that expected in dense photo-dissociation regions\cite{Rollig13}. The ratio is mostly constrained by the weak \ccc~emission on the west side of the disk and future higher sensitivity observations are needed to confirm this ratio. To test the sensitivity of parameters on the $^{12}$C/$^{13}$C ratio, we also run models with a fixed $^{12}$C/$^{13}$C ratio of 69. We find that the best-fitting parameters in these models are consistent with the results in Table~1 within uncertainties, except that the values of $\tau_0$ and $\delta$ increase by a factor of $\sim$2.

\subsection{Gas column density.}  With the radial profiles of temperature and line+dust optical depth from our parameterized model, we can derive the radial distribution of \ccc~column density by assuming that CO is in local thermal equilibrium (Supplementary Figure~\ref{fig:tau}). This assumption is reasonable since the gas number density at the mid-plane is expected to be much higher than the critical density of the CO $J$=3-2 line. To further convert the \ccc~column density distribution to a total gas mass distribution, we use the spatially unresolved HD (1-0) line flux to estimate the \ccc~abundance\cite{Bergin13}.   The HD (1-0) line has been used as a total mass tracer of the TW Hya disk since the D/H ratio is well calibrated as 1.5$\times$10$^{-5}$ in objects within $\sim$100\,pc of the Sun\cite{Linsky98}. With an upper state energy of 128.5\,K, the HD (1-0) line emission is dominantly from the warm region of the disk (T$>$20\,K), similar to the warm region where CO remains in the gas phase. Therefore, our \ccc~gas column density distribution makes a good approximation to the mass distribution of HD that contributes to the HD (1-0) emission, given $x_{\rm ^{13}C^{18}O}$ is a constant in the warm region throughout the disk. The HD (1-0) line flux is (6.3$\pm$0.7)$\times$10$^{-18}$ W m$^{-2}$ at 112\,$\mu$m\cite{Bergin13}. Using our \ccc~column density and temperature distribution, we find that 90\% of the HD (1-0) line flux comes from the inner 20\,AU of the disk, a region for which  the dust continuum emission at 112\,$\mu$m is optically thick ($\tau_{17{\rm AU}}\sim3.8$ for mm-sized dust grains), and therefore the line flux arises only from one side of the disk. Following the approach in Bergin et al. (2013)\cite{Bergin13}, we assume the HD (1-0) line is in LTE and scale the \ccc~column density to match the measured HD line flux. Propagating uncertainties in our best-fitting parameters and in the HD (1-0) line flux, we derive an average \ccc~abundance of 1.7$^{+1.3}_{-0.8}\times$10$^{-10}$ throughout the warm region of the TW Hya disk (T$>$20\,K).  This corresponds to a gas surface density profile of $\Sigma_g$ = 13$^{+8}_{-5}\times$(R/20.5\,AU)$^{-0.87^{+0.38}_{-0.26}}$\,g cm$^{-2}$ between 5-20.5\,AU.

\subsection{ The gas-to-dust mass ratio and its uncertainty sources.} Combining the uncertainties in our gas and dust surface density distribution, we derive a total gas-to- dust ratio of 140$^{+43}_{-31}$ for the 5-20.5\,AU region. The main uncertainty for the gas mass estimation is from the gas temperature because the HD (1-0) line emission has a strong temperature dependence\cite{Bergin13}. For the dust mass, we have not included the uncertainty of dust opacity due to dust growth. We use a $\kappa$ consistent with a dust population with an a$_{\rm max}$ of 1\,mm, but the $\kappa$ would decrease by a factor of two for an a$_{\rm max}$ of 1\,cm, and by a factor of 8 for 10\,cm. For this reason, we treat our estimation as a gas-to-mm-sized dust mass ratio. 

\subsection{Sanity check.} The critical assumption in the model above is that the \ccc, \cc, and dust continuum emission within the CO snowline can be characterized with the same temperature at each radius. Here we perform a sanity check to see if the assumption is valid in our best-fit model. The vertical distribution of gas and dust is characterized by their scale heights. Assuming a hydrostatic disk, we calculate the gas scale height using h$_g$ = $c_s/\Omega_K$, where $c_s$ is the disk sound speed at the mid-plane and $\Omega_K$ is the Keplerian angular velocity. The scale height of dust depends on turbulence level and the strength of the coupling between dust particles and turbulent gas\cite{Dubrulle95}. 
Assuming a medium $\alpha$ value of 10$^{-3}$, we find that the effective scale height of dust grains responsible for the 345\,GHz continuum emission is $\sim$40\% of that of the gas in the 5-21\,AU for our best-fit model. As shown in Supplementary Figure~\ref{fig:tau}, the \ccc, \cc~$J$=3-2 and dust optical depth is nearly flat inside 21\,AU and can be characterized with constant values for each species. Using a $h_d/h_g$ of 0.4 and three typical optical depths for the \ccc, \cc~and continuum emission, we show that the optically thin \ccc~and dust emission are from both sides of the disk in vertical regions within two gas scale heights (Figure~\ref{fig:cum_flux}). Although the optically thick \cc~line emission is mainly from one side of the disk,  80\% of its flux are from the region within two scale heights. Therefore the best-fit model is self-consistent in the treatment of the temperature. 

In more extreme cases that $\alpha$ approaches 10$^{-4}$, the mm-sized grains become significantly more settled than the gas\cite{Bai16}. To check the temperature difference between gas (micron-sized dust) and mm-sized dust grains, we run radiative transfer models for the TW Hya disk with two dust populations. Both populations have a power-law size distribution n(a)$\propto$a$^{-3.5}$ between a$_{\rm min}$ = 0.005\,${\rm \mu}$m and a$_{\rm max}$. The first population takes 10\% of the total dust mass, has an a$_{\rm max}$=1\,${\rm \mu}$m, and thus is vertically couple with the gas. The second population accounts for 90\% of the dust mass, has an a$_{\rm max}$=1\,mm, and has a scale height of 20\% of the gas scale height. 
We find that although the large grain population is cooler than the small grain population in the surface layer and outer disk region, the two populations share very similar temperature within one gas scale height between 5-20\,AU (Supplementary Figure~\ref{fig:temp_diff}). In summary, it is reasonable to assume that \ccc, \cc, and dust continuum emission within 20\,AU can be characterized with the same temperature.

\subsection{Data Availability Statement}

The data that support the plots within this paper and other findings of this study are available from the corresponding author upon reasonable request.

\end{methods}
{\Large \noindent Supplementary Materials}

\renewcommand{\tablename}{Supplementary Table}
\renewcommand{\figurename}{Supplementary Figure}

\begin{table}[hbp]
{\footnotesize
\caption{Fixed model parameters for the disk of TW Hya}
\begin{center}
\begin{tabular}{cccccccc}
\hline
\hline
Stellar mass & Distance & Inner radius & Outer radius& inclination &  & Turbulence & Position angle \\
\hline
M$_\star^{8}$& $d$\cite{vanLeeuwen07} & r$_{\rm in}$\cite{Rosenfeld12} & r$_{\rm out}$ & $i$\cite{Qi04} & V$_{\rm LSR}$\cite{Hughes11} & $v_{\rm turb}$ & PA\cite{Hughes11} \\
(M$_\odot$) & (pc) & (AU) & (AU) & (deg) & (km s$^{-1}$) & (km s$^{-1}$) & (deg)\\
\hline
0.6 & 55 & 2 &150 & 7 & 2.8 & 0.01 & 155\\
\hline
\end{tabular}
\end{center}
\label{tab:sup1}}

\end{table}

\begin{figure}[p!]
\centering
\includegraphics[width=4in]{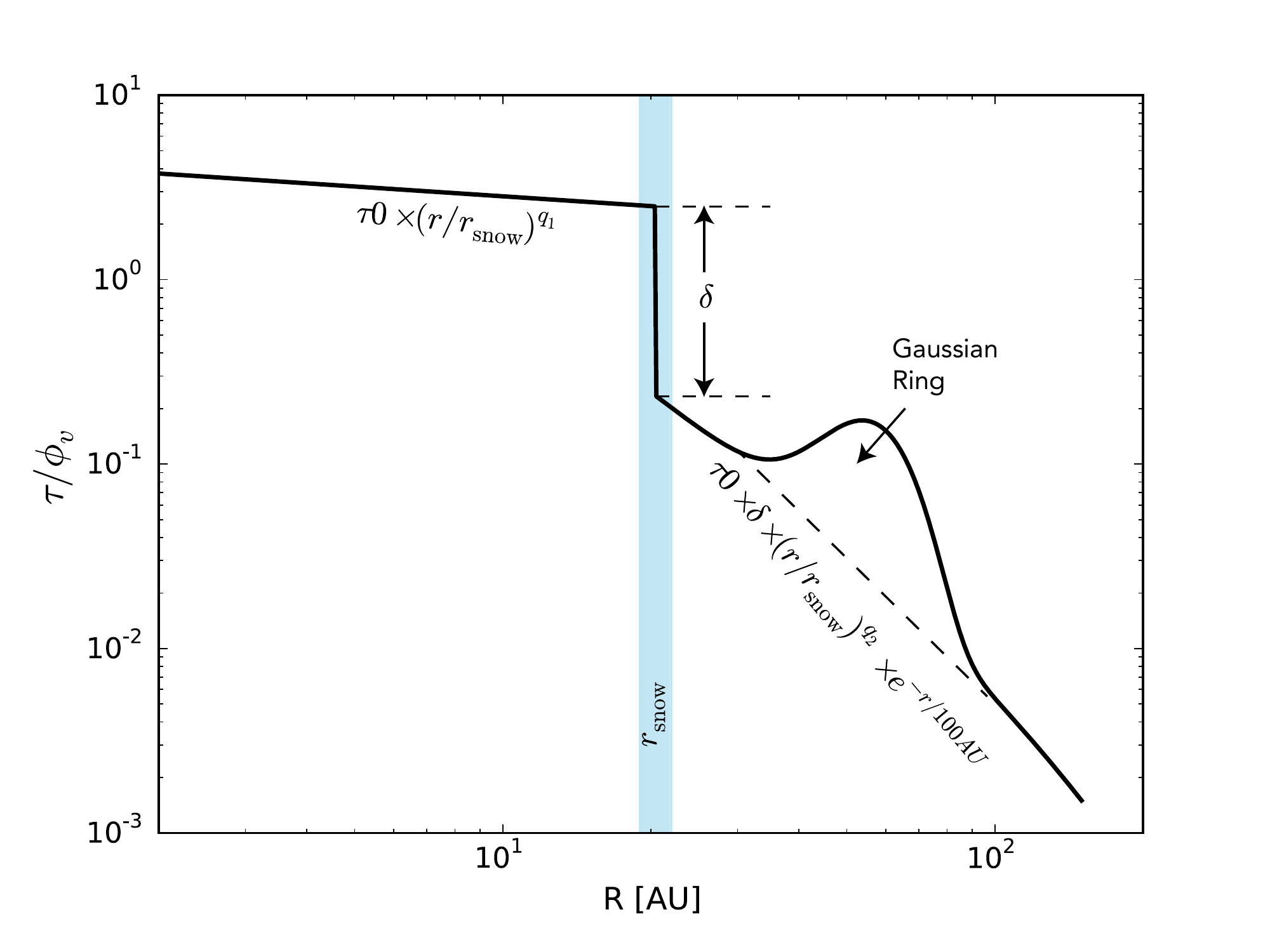}
\vspace{-0.45cm}
\caption{The radial distribution of the integrated line strength of the parameterized model. The model structure consist of two distinct radial zones: inside and outside the mid-plane CO snowline. The two zones are used to model the break of the CO gas column density distribution due to condensation. }
\label{fig:tau_model}
\vspace{-0.15cm}
\end{figure}

\clearpage

\begin{figure}[htbp]
\centering
\includegraphics[width=6in]{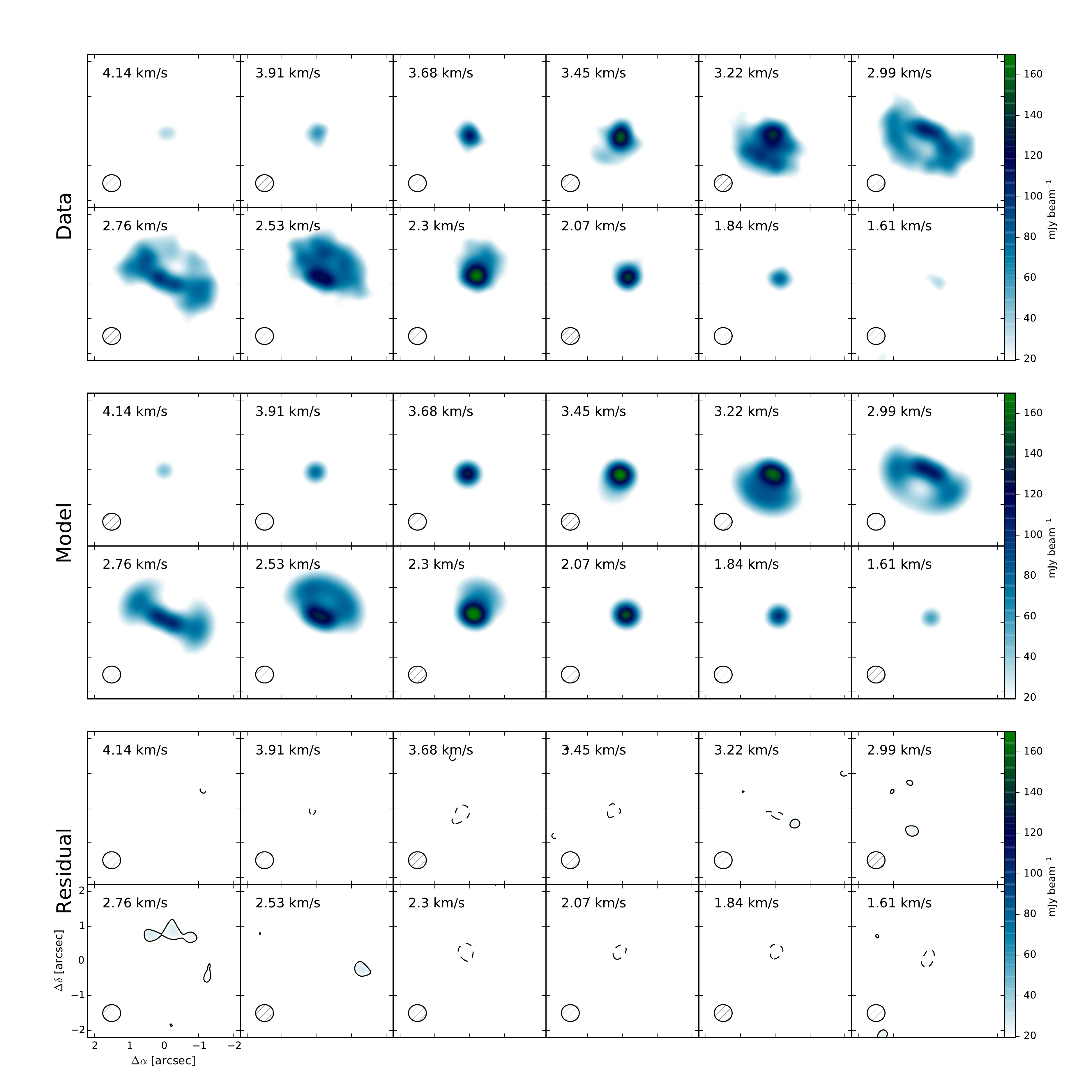}
\vspace{-0.6cm}
\caption{Direct comparison of the observed \cc~$J$=3-2 channel maps with synthetic data from the best-fit model in Table 1. The contour levels in the residual maps are set to be 3$\sigma$ (solid line) and -3$\sigma$ (dash line). }
\label{fig:cc}
\vspace{-0.15cm}
\end{figure} 

\begin{figure}[htbp]
\centering
\includegraphics[width=6in]{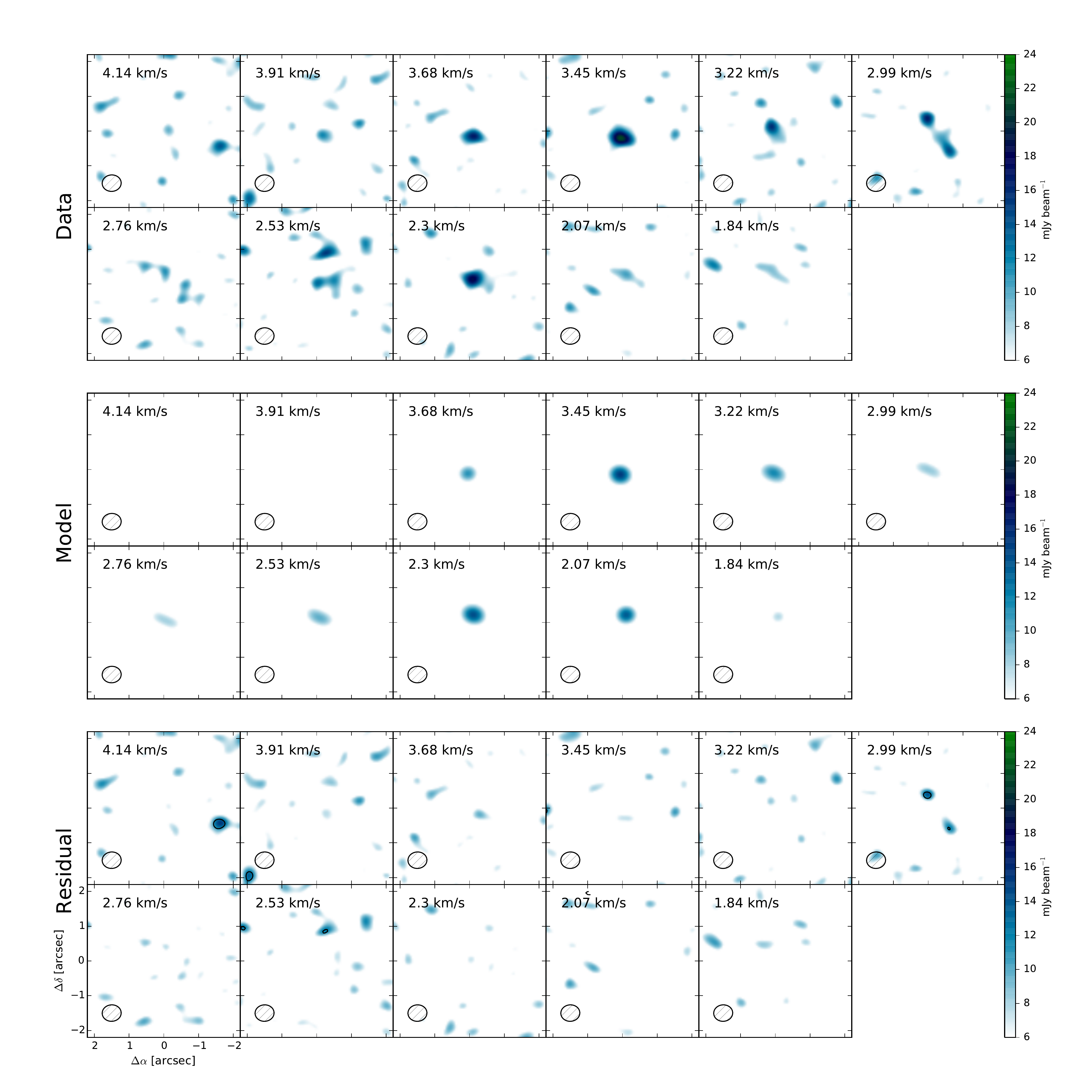}
\vspace{-0.6cm}
\caption{Direct comparison of the observed \ccc~$J$=3-2 channel maps with synthetic data from the best-fit model in Table 1.}
\label{fig:ccc}
\vspace{-0.15cm}
\end{figure}

\begin{figure}[p!]
\centering
\includegraphics[width=6in]{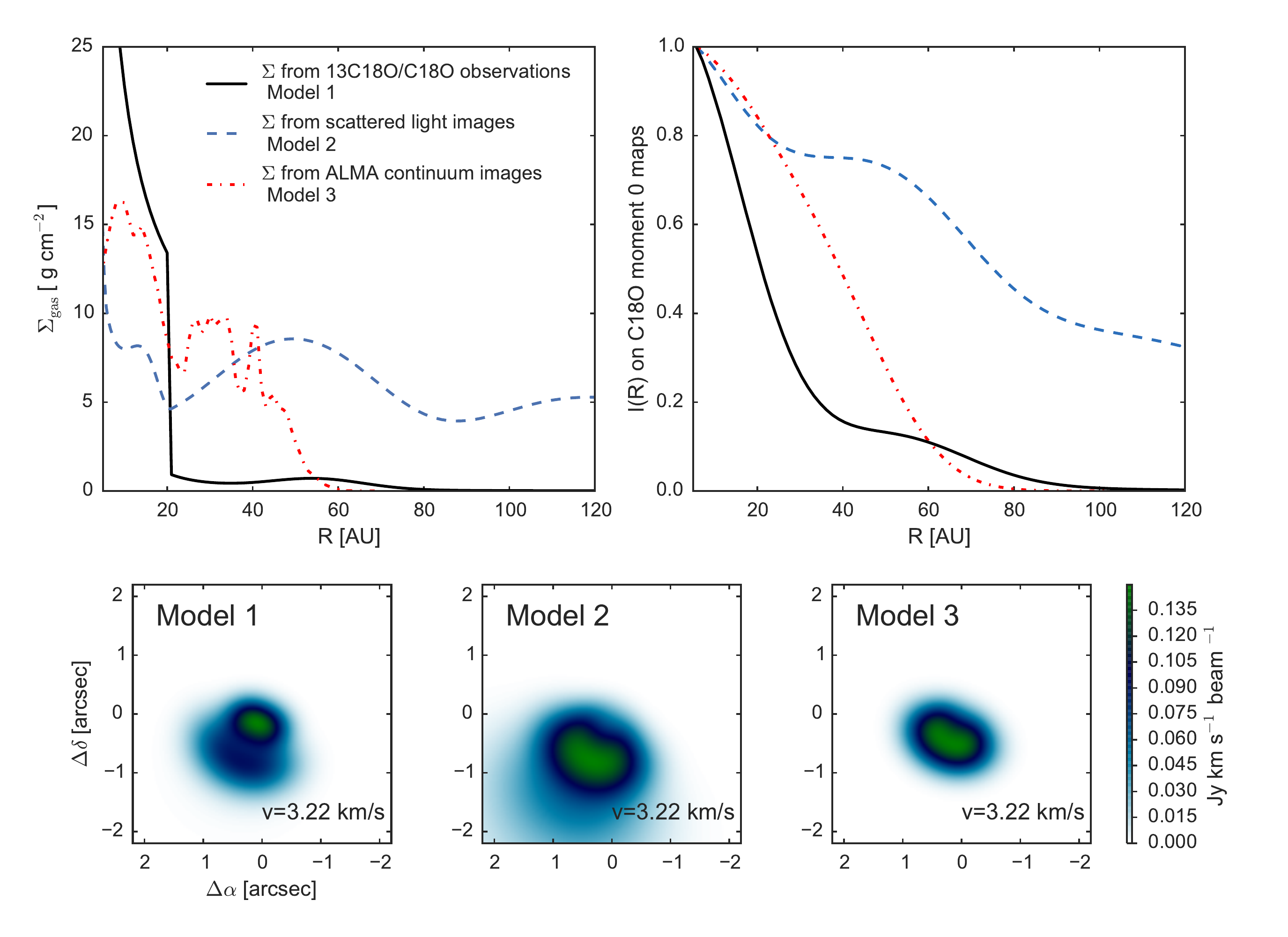}
\vspace{-0.45cm}
\caption{The \cc~(2-1) line emissions predicted by three models of the gas surface density distribution. Upper Left: The gas surface density profiles from three different models. Model 1 is the best-fitting model of this work, Model 2 is based on near Infrared scattered light images (ref.\,42), and Model 3 is derived from the 345\,GHz continuum emission. Upper Right: the radial profiles of the \cc\,(2-1) line moment zero map from the three models. It shows that the dust emission based models fail to reproduce the steep intensity decrease around 21\,AU in the \cc~observations. Lower panels: The \cc\,(2-1) line emission at the 3.22\,km s$^{-1}$ channel predicted by the three models. Only Model 1 can reproduce the narrow central peak observed in the \cc~line emissions (see Supplementary Figure~\ref{fig:cc}). The comparison of these models show that the sudden jump in the CO gas column density around 21\,AU is not seen in either small or large dust particle distributions and it is most likely caused by CO sublimation at the mid-plane.
}
\label{fig:f_model}
\vspace{-0.15cm}
\end{figure}

\begin{figure}[htbp]
\centering
\includegraphics[width=4in]{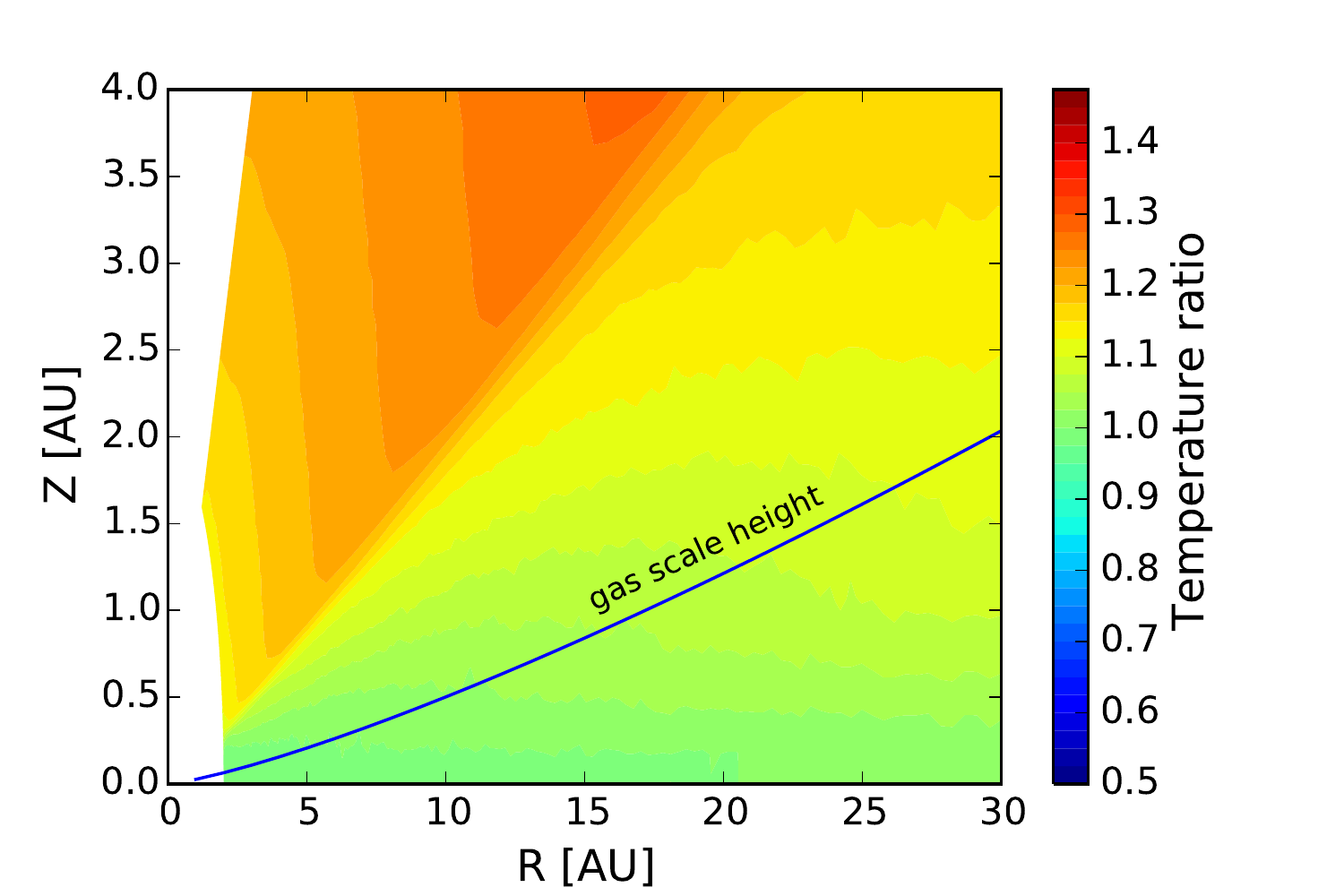}
\vspace{-0.6cm}
\caption{Map of the temperature ratio between a small and a large dust grain populations for a representative radiative transfer model of the TW Hya disk. The vertical scale height of the large dust grain population is only 20\% of that of the small dust population. }
\label{fig:temp_diff}
\vspace{-0.15cm}
\end{figure}

\end{document}